\documentclass[prb,twocolumn,superscriptaddress]{revtex4}

\usepackage{graphicx}
\usepackage{dcolumn}
\usepackage{amsmath}
\usepackage{color}

\begin{document}

\title{Electronic structure, quasiparticle renormalizations, and magnetic correlations in the alternating single-layer bilayer nickelate La$_5$Ni$_3$O$_{11}$}

\author{I. V. Leonov}
\affiliation{M. N. Mikheev Institute of Metal Physics, Russian Academy of Sciences, 620108 Yekaterinburg, Russia}
\affiliation{Institute of Physics and Technology, Ural Federal University, 620002 Yekaterinburg, Russia}

\begin{abstract}

Using the DFT+dynamical mean-field theory (DFT+DMFT) method, we study the normal-state electronic structure, orbital-dependent quasiparticle renormalizations, Fermi surface, and magnetic correlations in the recently discovered alternating single-layer bilayer hybrid Ruddlesden-Popper nickelate superconductor La$_5$Ni$_3$O$_{11}$ (1212-LNO).
Our results for the $\mathrm{DFT}+\mathrm{DMFT}$ electronic states exhibit qualitative differences for the structurally distinct single-layer and bilayer Ni ions, implying the importance of confinement and orbital-dependent correlations. The partially occupied Ni $e_g$ electronic states originating from the bilayer Ni ions form strongly renormalized quasiparticle bands with a large enhancement factor $m^*/m \sim 3.5$ and 4.2 for the Ni $x^2-y^2$ and $3z^2-r^2$ orbitals, respectively, implying the proximity of the Ni $3d$ states to orbital-dependent localization.
Moreover, the $e_g$ states of the single-layer Ni ions exhibit an orbital-selective Mott insulating state, with a narrow energy gap for the Ni $3z^2-r^2$ states and metallic, strongly incoherent (non-Fermi-liquid) Ni $x^2-y^2$ ones. 
Our analysis of magnetic correlations suggests the formation of intertwined spin and charge density wave stripes in the bilayer NiO$_6$ slab, in close similarity to the double-layer material. We identify two major instabilities: a leading one associated with a propagating wave vector ${\bf Q}=(\frac{1}{3},\frac{1}{3})$  ({associated, e.g., with an ``up-down-0" spin pattern or ``$d^8$--$d^8$--$d^8\underline{L}$'' charge-density wave with Ni $d^8\underline{L}$ ions bonded in an interlayer spin polaron}), competing with the bicollinear $(\frac{1}{4},\frac{1}{4})$ (``up-up-down-down") stripe state. In contrast, the single-layer Ni $3d$ electrons exhibit an instability toward the N\'eel-type magnetic state.
Under pressure, 1212-LNO undergoes an orbital-selective Mott insulator-to-metal phase transition, associated with metallization of the single-layer Ni $e_g$ states. As a result, the single-layer Ni $e_g$ bands exhibit non-Fermi-liquid (bad metal) behavior with strongly incoherent spectral weights near the Fermi level.
We note that correlation effects result in a remarkable reconstruction of the symmetry and strength of magnetic correlations as compared to those obtained within DFT. In fact, we observe a crossover from single-layer (as obtained within DFT) to double-layer dominated magnetic correlations.

\end{abstract}

\maketitle

%%%%%%%%%%%%%%%%%%%%%%%
\section{Introduction}

The recent discovery of unconventional superconductivity in the Ruddlesden-Popper (RP) nickelates La$_{n+1}$Ni$_n$O$_{3n+1-\delta}$ (LNO) with double-layer ($n = 2$ with a nominal ${2.5+}$ valence state of Ni ions) and trilayer ($n = 3$, Ni$^{2.67+}$) crystal structures has received much attention, placing this novel class of superconducting materials at the forefront of modern research \cite{Sun_2023,Wang_2024a,Hou_2023,YZhang_2024,WangWang_2024,Li_2025,Qiu_2025,
Ko_2025,Zhou_2025,Zhu_2024,Li_2024,Zhang_2025,
MZhang_2025,Pei_2026,WangYing_2025,MWang_2024}. It has been demonstrated that superconductivity appears at a high critical temperature of about 80-90~K in the high-quality single- and poly-crystalline bulk double-layer nickelates $R_3$Ni$_2$O$_7$ with $R=\mathrm{La}$, Pr, Sm under pressure above 15 GPa \cite{Sun_2023,Wang_2024a,Hou_2023,YZhang_2024,WangWang_2024,Li_2025,
Qiu_2025,Ko_2025,Zhou_2025,MWang_2024}. In the trilayer compound $T_c$ is sufficiently lower, about 20-40~K, appearing under similar pressure and compositions \cite{Zhu_2024,Li_2024,Zhang_2025,MZhang_2025,Pei_2026,WangYing_2025}. 

In both materials, superconductivity appears near a pressure-driven structural phase transition to a high-symmetry tetragonal crystal structure. The phase transition is accompanied by the suppression of a long-range spin-charge-density wave ordering. In fact, recent experiments show the presence of long-range magnetic correlations associated with a spin-charge density wave stripe ordering below about 140-150 K (at ambient pressure) \cite{Chen_2024a,Kakoi_2024,Agrestini_2024,Dan_2024,
Meng_2024,Khasanov_2024,Cao_2025,LiGong_2025,Dou_2026}. This structural transition suppresses the tiltings of the NiO$_6$ octahedra present at ambient pressure. As a result, the bandwidth of the partially occupied Ni $e_g$ orbitals increases, which leads to a reduction of the strength of electron-electron correlations {(with a crossover to a more itinerant behavior of the Ni $3d$ electrons)}. In addition, experiments suggest the appearance of strange metal behavior in the normal state of the double-layer and trilayer LNOs extending up to room temperature \cite{Sun_2023,Hou_2023,YZhang_2024,Wang_2024a}. This implies the crucial importance of the effects of electronic correlations and their complex interplay with the electronic structure properties \cite{Dong_2024,Yang_2024a,Liu_2024,Xie_2024,
MZhang_2024,Liu_2025}. 
In agreement with this, applications of the correlated electronic structure methods, e.g., using the $\mathrm{DFT}+\mathrm{dynamical}$ mean-field theory (DMFT) \cite{Georges_1996,Kotliar_2006} and $GW+\mathrm{DMFT}$ \cite{Biermann_2003,Tomczak_2017}  methods, show the crucial impact of orbital-dependent correlations, such as orbital-selective quasiparticle renormalizations and strong incoherence of the spectral weights associated with the Ni $e_g$ states near the Fermi level  \cite{Zhang_2023a,Shilenko_2023,Lechermann_2023,
Christiansson_2023,Liao_2023,Shen_2023b,Ryee_2024,
Craco_2024,Cao_2024,YYang_2023,Qin_2023,
Wang_2024,Leonov_2024a,LaBollita_2024a,Huang_2024,
Chen_2024b,Tian_2024b,Leonov_2026}.

{Moreover, the low-energy electronic structure of these materials exhibits strong covalency between the Ni $3d$ and O $2p$ states, implying the importance of negative (or small) charge-transfer energy effects. Indeed, the Ni ions adopt a nearly Ni$^{2+}$ state despite being nominally mixed-valence \cite{Shilenko_2023,Lechermann_2023,Leonov_2026,Leonov_2024a}.
This result is in agreement with recent x-ray absorption and resonant inelastic x-ray scattering experiments \cite{Li_2026,Chen_2026,Zhang_2026}, which demonstrate that interlayer hybridization and delocalization of the O $2p_z$ and Ni $3z^2-r^2$ orbitals, bound in an interlayer spin polaron \cite{Li_2026}, are crucial for superconductivity in bilayer nickelates.
}
While theoretical analysis of these systems converges toward s$^\pm$-type superconductivity (possibly competing with the $d$-wave component) \cite{ZhangLin_2024b,Yang_2024b,Zhang_2023c,Liu_2023b,Yang_2023,Heier_2024,Lu_2024,
Fan_2024,Sakakibara_2023a,Tian_2024a}, the microscopic origins of their anomalous normal state and superconductivity still remain a subject of active debate.

For the bilayer and trilayer LNOs, the $\mathrm{DFT}+\mathrm{DMFT}$ calculations reveal the complex interplay between the effects of electron correlations, Fermi surface nesting, and pressure. Most importantly, applications of the band structure and the $\mathrm{DFT}+\mathrm{DMFT}$ methods to study a low-pressure long-range magnetic phase suggest the emergence of double spin-charge-density wave ordering in the double-layer LNO under low pressure and temperature \cite{Leonov_2025,LaBollita_2024b,BZhang_2024,Ni_2024,Tian_2025}, in agreement with recent experiments \cite{Chen_2024a,Kakoi_2024,Agrestini_2024,Dan_2024,
Meng_2024,Khasanov_2024,Cao_2025,LiGong_2025,Dou_2026}. 
Overall, this suggests that spin and charge stripe fluctuations are important for understanding the mechanism of superconductivity in the layered RP nickelates.

The layered RP nickelates are found to exhibit unusual structural complexity associated with different stacking of the perovskite-type (LaNiO$_3$)$_n$ structural blocks separated by the NaCl-type LaO layers (akin to van der Waals-type heterostructuring), e.g., as in the experimentally discovered single-layer trilayer (``1313'') structural polymorph of the bilayer La$_3$Ni$_2$O$_7$  (with a uniform Ni$^{2.5+}$ state) \cite{Chen_2024,LiGuo_2024,Puphal_2024,WangChen_2024,
Abadi_2024,Au-Yeung_2025}. The latter has the uniform bilayer-by-bilayer stacking (2222-LNO). This approach allows for a more precise tuning of the properties by stacking atomic layers with distinct electronic and magnetic properties. Interestingly, the hybrid single-layer trilayer polymorph becomes superconducting under pressure with a sufficiently reduced critical temperature of $\sim$4~K than that in the conventional trilayer system ($\sim$20-40~K) \cite{Huang_2025}. Very recently, superconductivity with a high critical temperature $\sim$64~K has been reported in single crystals of the single-layer bilayer (``1212'') hybrid RP nickelate La$_5$Ni$_3$O$_{11}$ (with a nominal Ni$^{2.33+}$ state) under pressure above $\sim$15 GPa \cite{ShiPeng_2025}. The latter shows an alternative stacking of the monolayer La$_2$NiO$_4$ (with Ni$^{2+}$ ions) and bilayer LNO (with Ni$^{2.5+}$ ions) slabs along the $c$ axis. We note that the bulk monolayer La$_2$NiO$_{4+\delta}$ with a nominal Ni$^{2+}$ state has an orthorhombic structure with the rotated NiO$_6$ octahedra. It exhibits a N\'eel-type antiferromagnetic Mott insulating state close to its stoichiometric composition at low temperatures \cite{Yamada_1992,Fabbris_2017,Petsch_2023}.

In spite of recent active research, certain challenges remain unresolved, e.g., those concerning the impact of electronic correlations on the electronic structure, magnetism, and superconductivity in the hybrid 1212 and 1313 LNO phases of nickelates \cite{Lechermann_2024,LaBollita_2024,ZhangLin_2024,
Ouyang_2025,Sharma_2026,ZhangLin_2025,ZhangChen_2026,LaBollita_2026}. Applications of the $\mathrm{DFT}+\mathrm{DMFT}$ electronic structure calculations suggest a key role of Mott localization of the Ni $e_g$ electrons for the single layer, while the low-energy physics is dominated by renormalized quasiparticle Ni $e_g$ bands associated with the Ni ions in the bilayer or trilayer slabs at ambient and at high pressures \cite{Lechermann_2024,LaBollita_2024,LaBollita_2026,
Ouyang_2025}. This picture is consistent with recent angle-resolved photoemission spectroscopy measurements of 1313-LNO at ambient pressure \cite{Au-Yeung_2025,Abadi_2024}. On the other hand, previous constrained random phase approximation calculations based on the DFT band structure show that the leading instability in both 1212- and 1313-LNO is dominated by the monolayer NiO$_6$ block \cite{ZhangLin_2024,ZhangChen_2026,ZhangLin_2025}. In fact, the impact of electron correlations on the electronic, magnetic, and superconducting properties of the hybrid 1212- and 1313-LNO is still a subject of debate.

We address this topic in our present work, which is motivated by the recent reports on superconductivity in the highly pressurized 1212-LNO (La$_5$Ni$_3$O$_{11}$) \cite{ShiPeng_2025}. In our study, we explore the effects of correlations on the normal-state electronic structure and magnetic properties of 1212-LNO under pressure. Using the $\mathrm{DFT}+\mathrm{DMFT}$ approach, we explore the correlated electronic structure, orbital-dependent quasiparticle renormalizations, incoherence of the low-energy spectral weights, Fermi surface topology, and magnetic correlations of paramagnetic 1212-LNO. 
Our results reveal remarkable layer- and orbital-dependent correlation effects of the Ni $3d$ states. In agreement with previous studies \cite{LaBollita_2026}, we find that the low-energy electronic states are dominated by the strongly renormalized quasiparticle Ni $e_g$ bands associated with Ni ions in the bilayer NiO$_6$ block. For the monolayer Ni ions, we obtain an orbital-selective Mott insulating state with a narrow energy gap for the Ni $3z^2-r^2$ orbital states and non-Fermi-liquid behavior for the Ni $x^2-y^2$ states. The latter contribute strongly incoherent spectral weights near $E_F$. 
This result suggests the crucial importance of layer- and orbital-dependent Mott localization in 1212-LNO. Moreover, our analysis of the {\bf k}-resolved spectral functions and correlated Fermi surfaces shows significant incoherence of the Ni $3z^2-r^2$ states associated with the bilayer NiO$_6$ block, implying the proximity of the bilayer Ni $e_g$ states to orbital-selective localization.  

Our results propose the emergence of spin and charge density wave stripes (at low pressure and temperature), which seem to be important for understanding the anomalous properties of 1212-LNO, similarly to the conventional bilayer and trilayer LNOs. A leading instability is associated with double spin-charge density wave stripe ordering in the bilayer NiO$_6$ block with an in-plane propagating wave vector ${\bf Q}=(\frac{1}{3},\frac{1}{3})$. {This behavior can be interpreted in terms of two possible scenarios. The first is an instability toward an ``up-down-0'' spin-density wave.  The second is associated with the ``$d^8$--$d^8$--$d^8\underline{L}$'' charge-density wave state with the Ni $d^8\underline{L}$ ions bound in an interlayer spin polaron \cite{Li_2026}. This suggests a 
``up-down-UP-up-down-DOWN'' spin-density wave ordering (where up/down are the up/down spins of the Ni$^{2+}$ $3d^8$ ions; UP/DOWN are those of the Ni $3d^8\underline{L}$ ions).}
{The ${\bf Q}=(\frac{1}{3},\frac{1}{3})$ state} is found to strongly compete with a ${\bf Q}=(\frac{1}{4},\frac{1}{4})$ bicollinear stripe state (with  an {``up-up-down-down''} spin pattern). The Ni $3d$ electrons in the single-layer exhibit the N\'eel-type magnetic instability.

We observe a striking reconstruction of the magnetic excitation spectrum caused by correlation effects, highlighting the crucial importance of strong correlations. It leads to a change of the leading instability to that dominated by the bilayer NiO$_6$ block, in close similarity to the double-layer 2222-LNO. This is in sharp contrast to the regime of single-layer dominated instability predicted within DFT. Under pressure, 1212-LNO undergoes an orbital-selective
Mott insulator-to-metal phase transition, associated with metallization of the monolayer Ni $e_g$ states. For the monolayer Ni ions the $x^2-y^2$ and $3z^2-r^2$ orbital self-energies show strongly non-Fermi-liquid behavior. Overall, the strength of correlation effects is seen to be sufficiently stronger for the monolayer than for the bilayer Ni $e_g$ states. This implies the importance of confinement and multiorbital Coulomb correlations to explain the electronic and magnetic properties of 1212-LNO.

%%%%%%%%%%%%%%%%%%%%%%%

\section{Computational details}

In this work, using a fully charge self-consistent $\mathrm{DFT}+\mathrm{DMFT}$ method\cite{Georges_1996,Kotliar_2006} implemented with plane-wave pseudopotentials \cite{Leonov_2020a,Leonov_2024b}, we study the normal-state electronic structure, quasiparticle renormalizations, Fermi surface, and magnetic correlations in the hybrid Ruddlesden-Popper 1212-LNO. Our $\mathrm{DFT}+\mathrm{DMFT}$ calculations are performed in the paramagnetic (PM) state. In our calculations for the low pressure phase, we use the experimentally determined $Cmmm$ structure with the lattice parameters $a=5.426$ \AA, $b=5.448$ \AA, and $c=16.575$~\AA. For the high-pressure phase, we adopt the $P4/mmm$ structure with the lattice parameters $a=5.278$ \AA, and $c=16.085$~\AA\ taken at $\sim$20 GPa  and $a=5.211$ \AA, and $c=15.876$~\AA\ at $\sim$30 GPa  \cite{ShiPeng_2025}. The atomic coordinates are relaxed using the nonmagnetic DFT. In DFT, we employ the generalized gradient approximation with the Perdew-Burke-Ernzerhof (PBE) exchange functional as implemented in the Quantum ESPRESSO package \cite{Giannozzi_2009,Giannozzi_2017,DalCorso_2014}. In our $\mathrm{DFT}+\mathrm{DMFT}$ calculations, we explicitly include the Ni $3d$, La $5d$, and O $2p$ valence states, by constructing a basis set of atomic-centered Wannier functions within the energy window spanned by these bands \cite{Anisimov_2005,Marzari_2012}. This allows us to take into account charge transfer effects between the partially occupied Ni $3d$, La $5d$, and O $2p$ states, accompanied by the strong electron-electron interactions in the Ni $3d$ shell.

In DMFT, the realistic many-body problem is solved using the continuous-time hybridization expansion (segment) quantum Monte Carlo (CTQMC) method \cite{Gull_2011} with improved estimators for the self-energy \cite{Boehnke_2011,Hafermann_2012}. In CTQMC, we solve two distinct impurity problems for the structurally distinct Ni ions (in the monolayer and in the bilayer NiO$_6$ blocks). The effects of electron correlations in the Ni $3d$ shell are treated by using the onsite Hubbard interaction $U = 6$ eV and Hund's exchange $J = 0.95$ eV, taken in accordance with previous $\mathrm{DFT}+\mathrm{DMFT}$ calculations of the RP nickelates \cite{Lechermann_2023,Lechermann_2024,Shilenko_2023,Leonov_2024a,Leonov_2025,Leonov_2026}. The La $5d$ and O $2p$ valence states are uncorrelated and are treated on the DFT level within the self-consistent $\mathrm{DFT}+\mathrm{DMFT}$ approach. In our calculations, we use the fully localized double-counting correction evaluated from the self-consistently determined local occupations. We use  Pad\'e approximants for analytic continuation of the self-energy from the Matsubara frequencies [$\Sigma(i\omega_n)$] to the real energy axis [$\Sigma(\omega)$] in order to evaluate the {\bf k}-resolved spectral functions and Fermi surfaces.

\section{Results and discussion}
{
\subsection{Electronic structure}}

\begin{figure}
\centerline{\includegraphics[width=0.5\textwidth,clip=true]{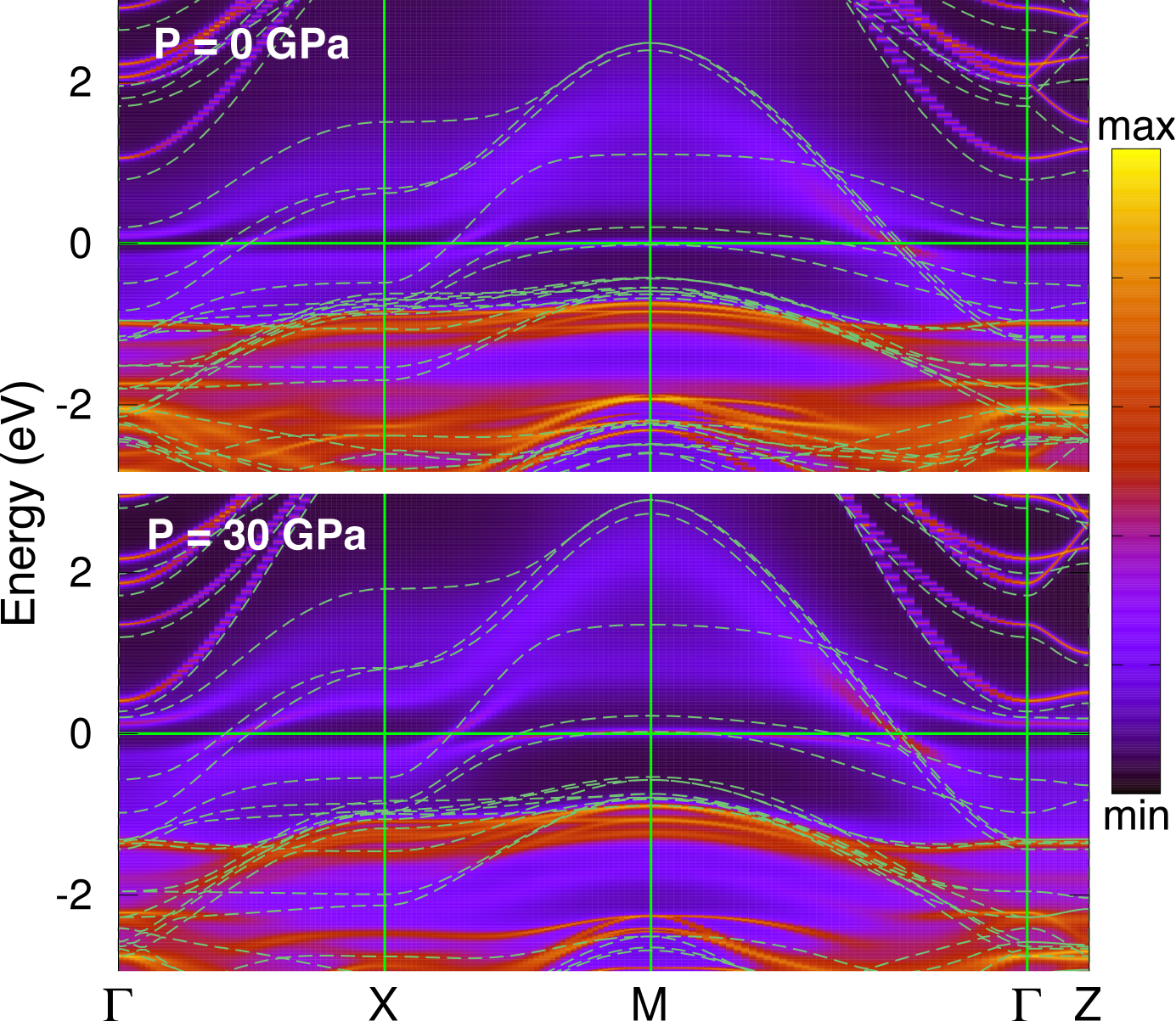}}
\caption{{\bf k}-resolved spectral functions of PM 1212-LNO calculated by $\mathrm{DFT}+\mathrm{DMFT}$ with $U=6$ eV and $J=0.95$ eV at $T = 298$~K. 
The calculations are performed for the low-pressure $Cmmm$ and high-pressure $P4/mmm$ crystal structures using experimental lattice parameters and optimized atomic positions. Our results are compared with the nonmagnetic DFT calculations (shown with dashed green lines).
}
\label{Fig_1}
\end{figure}

\begin{figure}
\centerline{\includegraphics[width=0.5\textwidth,clip=true]{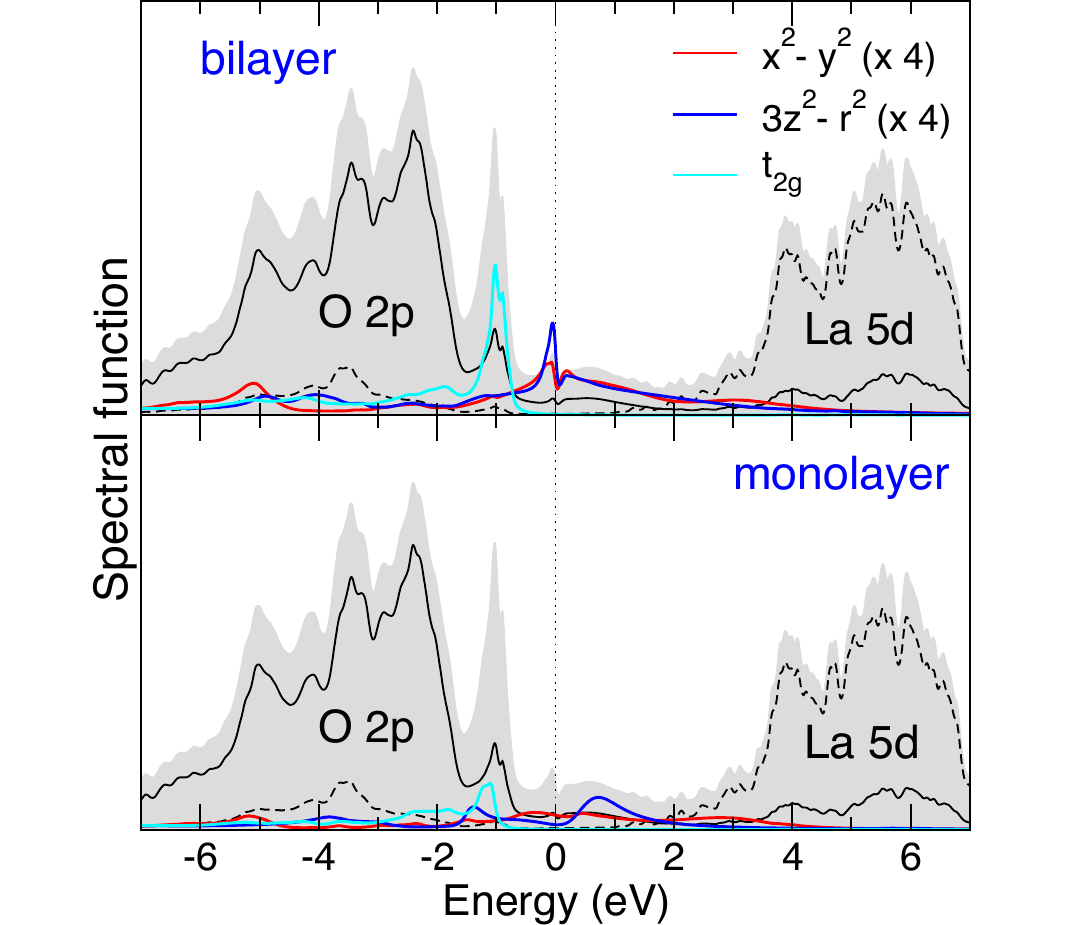}}
\caption{
Our results for the layer-resolved orbital-dependent spectral functions of 1212-LNO at low pressure calculated by $\mathrm{DFT}+\mathrm{DMFT}$ at $T = 298$~K. The partial Ni $t_{2g}$, $x^2-y^2$, and $3z^2-r^2$ orbital contributions are shown (per orbital). Ni $x^2-y^2$, and $3z^2-r^2$ orbital states are magnified for better readability.
}
\label{Fig_2}
\end{figure}

\begin{figure}
\centerline{\includegraphics[width=0.5\textwidth,clip=true]{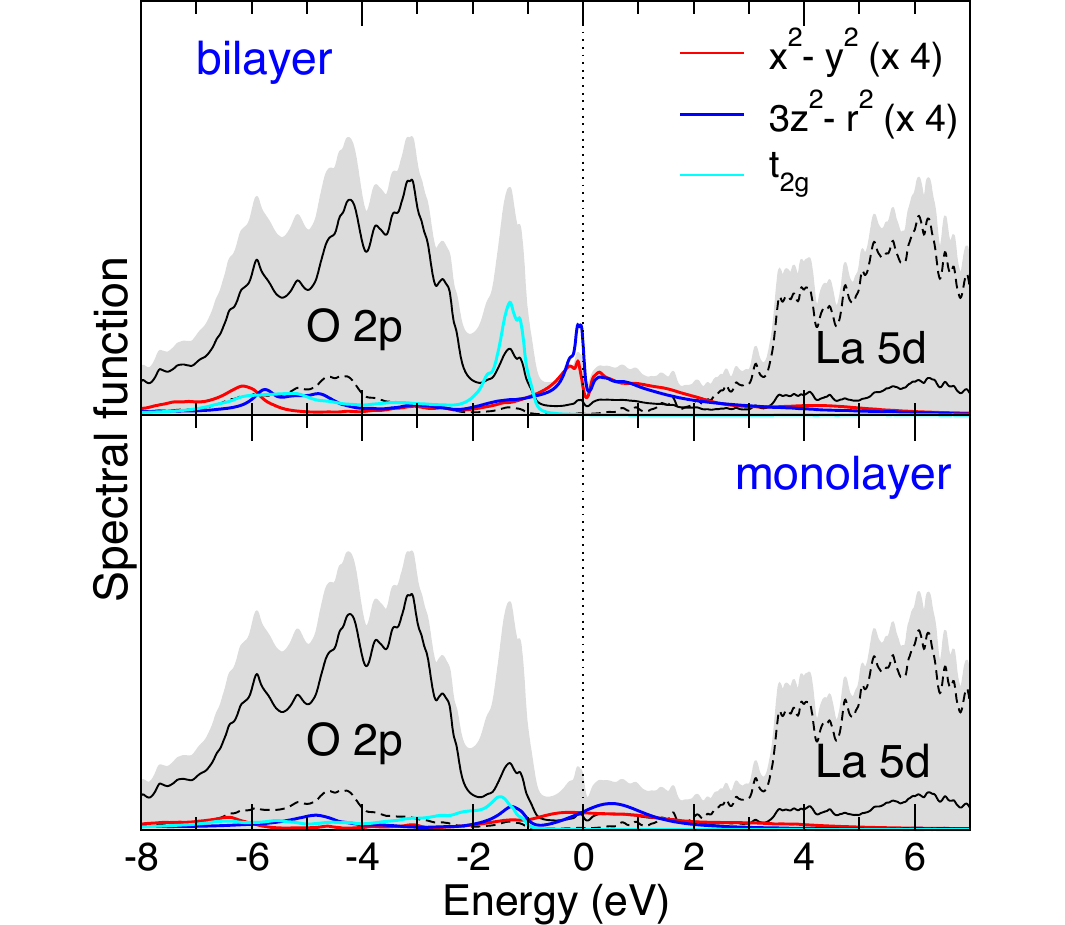}}
\caption{Orbital-dependent spectral functions of the high-pressure $P4/mmm$ phase of 1212-LNO calculated by $\mathrm{DFT}+\mathrm{DMFT}$ under pressure $\sim$30 GPa at $T = 298$~K. 
}
\label{Fig_3}
\end{figure}

Using $\mathrm{DFT}+\mathrm{DMFT}$ we study the effects of pressure and onsite Coulomb correlations on the normal-state electronic structure and magnetic properties of PM 1212-LNO at room temperature. In our calculations, we employ the experimentally determined $Cmmm$ crystal structure for the low-pressure 1212-LNO \cite{ShiPeng_2025} with relaxed atomic coordinates (within the nonmagnetic DFT). In Fig.~\ref{Fig_1}, we display the {\bf k}-resolved spectral functions of 1212-LNO calculated using the $\mathrm{DFT}+\mathrm{DMFT}$ method in comparison to the nonmagnetic DFT band structure. The partial Ni $3d$, La $5d$, and O $2p$ spectral functions are shown in Figs.~\ref{Fig_2} and \ref{Fig_3}, respectively. We compare our results with those obtained using $\mathrm{DFT}+\mathrm{DMFT}$ for the high-pressure $P4/mmm$ crystal structure taken at $\sim$30 GPa \cite{ShiPeng_2025} (for the experimental lattice parameters and relaxed atomic positions). Our DFT+DMFT calculations of the high-pressure 1212-LNO with the lattice structure parameters taken at $\sim$20 GPa yield similar results.

Our results for the {\bf k}-resolved spectra show a remarkable bonding-antibonding splitting of the Ni $e_g$ bands near the Fermi level associated with strong interlayer coupling of the Ni $e_g$ orbitals in the bilayer NiO$_6$ block (at both ambient and high pressure). 
This result corroborates previously obtained results for the double-layer 2222- and trilayer 3333-LNO phases. We find that the electronic states in the energy window of $-2$ to 2 eV near the Fermi level are dominated by the Ni $x^2-y^2$ and $3z^2-r^2$ orbitals. The Ni $t_{2g}$ states are fully occupied and are located between $-2$ to $-1$ eV below the Fermi level. The occupied O $2p$ orbitals experience strong hybridization with the Ni $e_g$ and La $5d$ states and appear at about $-4$ eV below $E_F$. The rare-earth element La $5d$ orbitals are empty and are located above 1 eV. 

Our $\mathrm{DFT}+\mathrm{DMFT}$ results exhibit the crucial importance of strong electron-electron interactions. Indeed, the partially occupied Ni $e_g$ states exhibit significant quasiparticle renormalizations near the Fermi level concomitant with substantial orbital-selective incoherence of the spectral weights (bad metal behavior). In the {\bf k}-resolved spectra, nearly flat-band dispersions emerge at $\sim$100 meV below and above the Fermi level near the Brilloun zone $X$ point. This behavior is associated with the planar Ni $x^2-y^2$ orbital states, indicating the emergence of flat-band physics because of a low-dimensional Van Hove singularity in the electronic spectra near $E_F$. In comparison to the bare DFT results, the latter is seen to be shifted toward the Fermi level because of strong quasiparticle renormalizations caused by correlation effects.

The self-consistently evaluated Ni $x^2-y^2$ and $3z^2-r^2$ Wannier orbital occupations are close to half filling ($\sim$0.56 and 0.6 per spin orbit for the Ni $x^2-y^2$ and $3z^2-r^2$ orbitals for the mono- and are 0.56 and 0.61 for the bilayer Ni ions, respectively). This suggests that the orbital polarization between the Ni $x^2-y^2$ and $3z^2-r^2$ orbitals is rather weak. Moreover, the calculated total Ni $3d$ Wannier orbital occupations for the mono- and bilayer Ni ions are nearly the same, about 8.29 and 8.28, respectively. {That is, the} Ni ions adopt a nearly 2+ valence state, which sufficiently differs from the nominal (uniform) oxidation state Ni$^{2.33+}$ or the interlayer valence skipping with Ni$^{2+}$ for the monolayer and Ni$^{2.5+}$ for the bilayer Ni ions. 
{In agreement with this}, our estimate of the weights of different atomic configurations of the Ni $3d$ electrons being fluctuating between various valence configurations within
DMFT gives 0.6 (0.56) and 0.32 (0.34) for the Ni $d^8$ and $d^9\underline{L}$ configurations for the monolayer (bilayer) Ni ions, respectively. 

{Our results highlight} the importance of negative charge transfer \cite{Zaanen_1985} between the Ni $3d$ and O $2p$ states, in agreement with previously obtained results for the double- and trilayer systems \cite{Shilenko_2023,Lechermann_2023,Leonov_2026,Leonov_2024a}.
{In fact, the low-energy structure of 1212-LNO can be characterized as that in a negative (or small) charge-transfer energy regime with a (nearly) Ni $3d^8$ configuration with significant oxygen ligand hole character, in agreement with strong hybridization between the Ni $3d$ and O $2p$ states. 
Our analysis of the O $2p$ Wannier orbital occupations shows a nearly uniform hole density distribution over oxygens. We find that the O $2p$ occupancies are within 2.59-2.63 (per spin). For all oxygen atoms, except the interbilayer apical oxygen, the O $2p$ Wannier orbital occupancies are nearly the same, implying 
that orbital polarization is negligible. 
In contrast to this, the interbilayer apical oxygens are characterized by a hole primarily residing at the O $2p_z$ orbital. The calculated O $2p_z$ and $2p_{x/y}$ orbital occupations are $\sim$0.82 and 0.89 per spin orbit. For comparison, for the planar oxygens bridging Ni ions in the \emph{ab} plane 
the difference of the O $2p_z$ and $2p_{x/y}$ orbital occupations is less 0.015; it is below than 0.002 for the outer apical oxygen.
We note that the O $2p_z$ orbital occupations of the interbilayer apical oxygens are nearly the same in the ambient $Cmmm$ and high-pressure $P4/mmm$ phases (decrease by $\sim$0.002 upon compression to 30 GPa).
It is worth noting that polarization of the inter-bilayer apical O $2p$ orbitals is larger than that for the Ni $e_g$ states. This result is in agreement with the recent x-ray absorption and resonant inelastic x-ray scattering experiments \cite{Zhang_2026,Chen_2026,Li_2026}. The x-ray spectroscopy experiments show that superconductivity in 2222-LNO thin films is primarily associated with delocalization of the O $2p_z$ and Ni $3z^2-r^2$ orbitals. 
In fact, the resulting orbital polarization of the O $2p$ states characterized by a hole residing in the O $2p_z$ orbital bridging the Ni $3d^8\underline{L}$ ions closely resembles 
the interlayer five-spin polaron state with S=3/2 proposed by J. Li \emph{et al.} \cite{Li_2026}.

}

{
\subsection{Site- and orbital-selective correlations}}

Our DFT results show remarkably different Ni $e_g$ ``noninteracting'' bandwidths for the mono- and bilayer Ni ions. The latter are about 3.5 and 2 eV for the planar $x^2-y^2$ and $3z^2-r^2$ orbitals for the monolayer Ni ions. For the bilayer Ni ions, the Ni $x^2-y^2$ and $3z^2-r^2$ orbital bandwidths are significantly larger, about 3.9 and 2.8 eV, respectively. This suggests a strong layer and orbital dependence of correlation effects in 1212-LNO. In fact, we obtain that the $\mathrm{DFT}+\mathrm{DMFT}$ electronic states of the Ni ions in the mono- and bilayer NiO$_6$ blocks exhibit \emph{qualitative} differences. The partially occupied Ni $x^2-y^2$ and $3z^2-r^2$ electronic states originating from the bilayer Ni ions form quasiparticle bands between $-2$ and 2 eV near $E_F$, with a quasiparticle peak at the Fermi level. In contrast, the $e_g$ states originating from the monolayer Ni ions show an orbital-selective Mott insulating state, in agreement with {previous studies}~\cite{LaBollita_2026}. Thus, the Ni $3z^2-r^2$ states show a narrow energy gap with the pronounced lower- and upper Hubbard subbands located at about $-1.2$ eV below and 0.8 eV above the Fermi level. The monolayer Ni $x^2-y^2$ states are metallic and show strongly incoherent spectral weight distributed near $E_F$ (exhibiting bad metal behavior).

\begin{figure}
\centerline{\includegraphics[width=0.5\textwidth,clip=true]{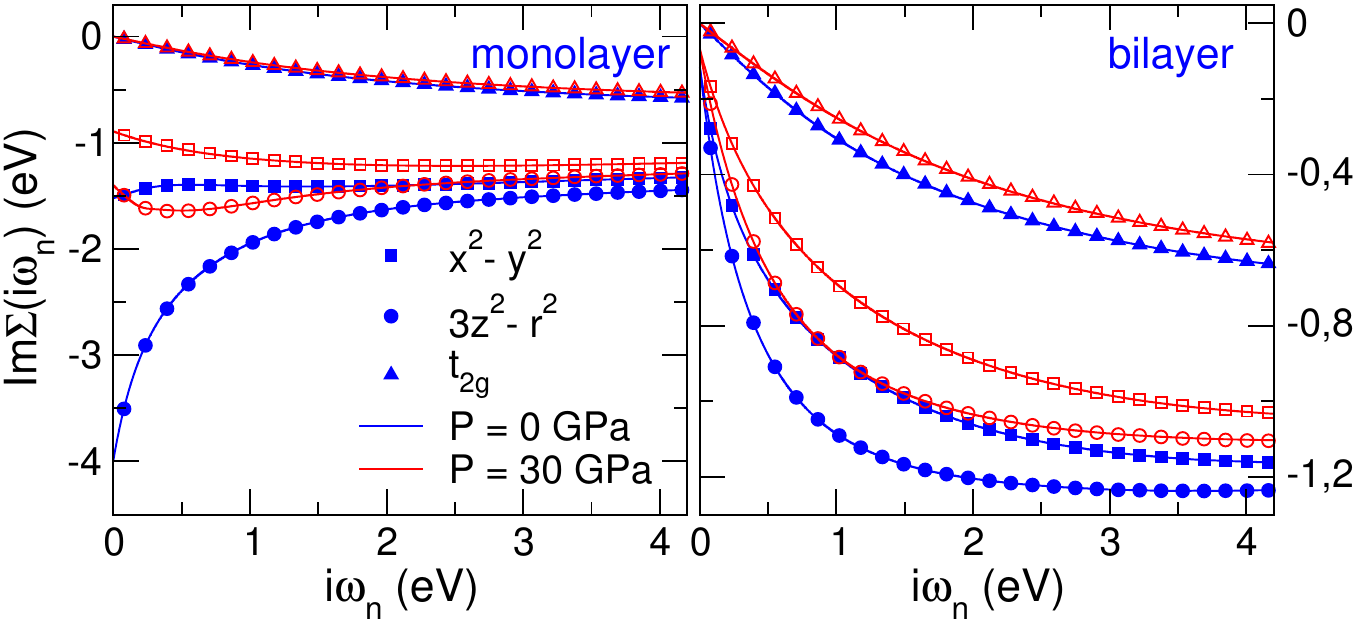}}
\caption{
Orbital-dependent Ni $3d$ self-energies $\mathrm{Im}[\Sigma(i\omega_n)]$ on the
Matsubara axis for the crystallographically distinct Ni sites in the mono- (left) and bilayer NiO$_6$ blocks (right) for different pressures (shown with symbols). Our results for the Pad\'e interpolation of $\mathrm{Im}[\Sigma(i\omega_n)]$ are depicted with lines.
}
\label{Fig_4}
\end{figure}

This result is in line with our analysis of the Ni $3d$ self-energies on the Matsubara contour. For the bilayer Ni ions, the Ni $3d$ self-energies show a typical Fermi-liquid-like behavior with remarkable renormalizations and quasiparticle damping of the Ni $e_g$ states (see Fig.~\ref{Fig_4}). Indeed, our estimate of the enhancement factors $m^*/m$ for the Ni $3d$ orbitals evaluated as $m^*/m = [1 - \partial \mathrm{Im}[\Sigma(i\omega)]/\partial i\omega]|_{i\omega \rightarrow 0}$ gives relatively large band renormalizations of about 3.5 and 4.2 for the bilayer Ni $x^2-y^2$ and $3z^2-r^2$ orbitals. For the $3z^2-r^2$ states, the quasiparticle damping evaluated as $\mathrm{Im}[\Sigma(i\omega_n)]$ at the first Matsubara frequency is noticeably higher, about 0.33 eV, compared with 0.28 eV for the $x^2-y^2$ states. Using Pad\'e extrapolation for the self-energy, we find about 0.12 for $\omega=0$ for the bilayer Ni $x^2-y^2$ and $3z^2-r^2$ orbitals, respectively. 

For the monolayer Ni ions, the  $x^2-y^2$ and $3z^2-r^2$ orbital self-energies show pronounced non-Fermi-liquid behavior with the imaginary part of the self-energy $\mathrm{Im}[\Sigma(i\omega_n)]$ diverging at the lowest Matsubara frequencies. This divergence is more pronounced for the Ni $3z^2-r^2$ orbitals, associate with the formation of an orbital-selective Mott insulating state. At the same time, the fully occupied Ni $t_{2g}$ states are sufficiently coherent, with $\mathrm{Im}[\Sigma(i\omega_n)]$ below 0.03 eV at the first Matsubara frequency for both the mono- and bilayer Ni ions. The corresponding enhancement factor for the Ni $t_{2g}$ orbitals is about $\sim 1.35$.

\begin{figure}
\centerline{\includegraphics[width=0.5\textwidth,clip=true]{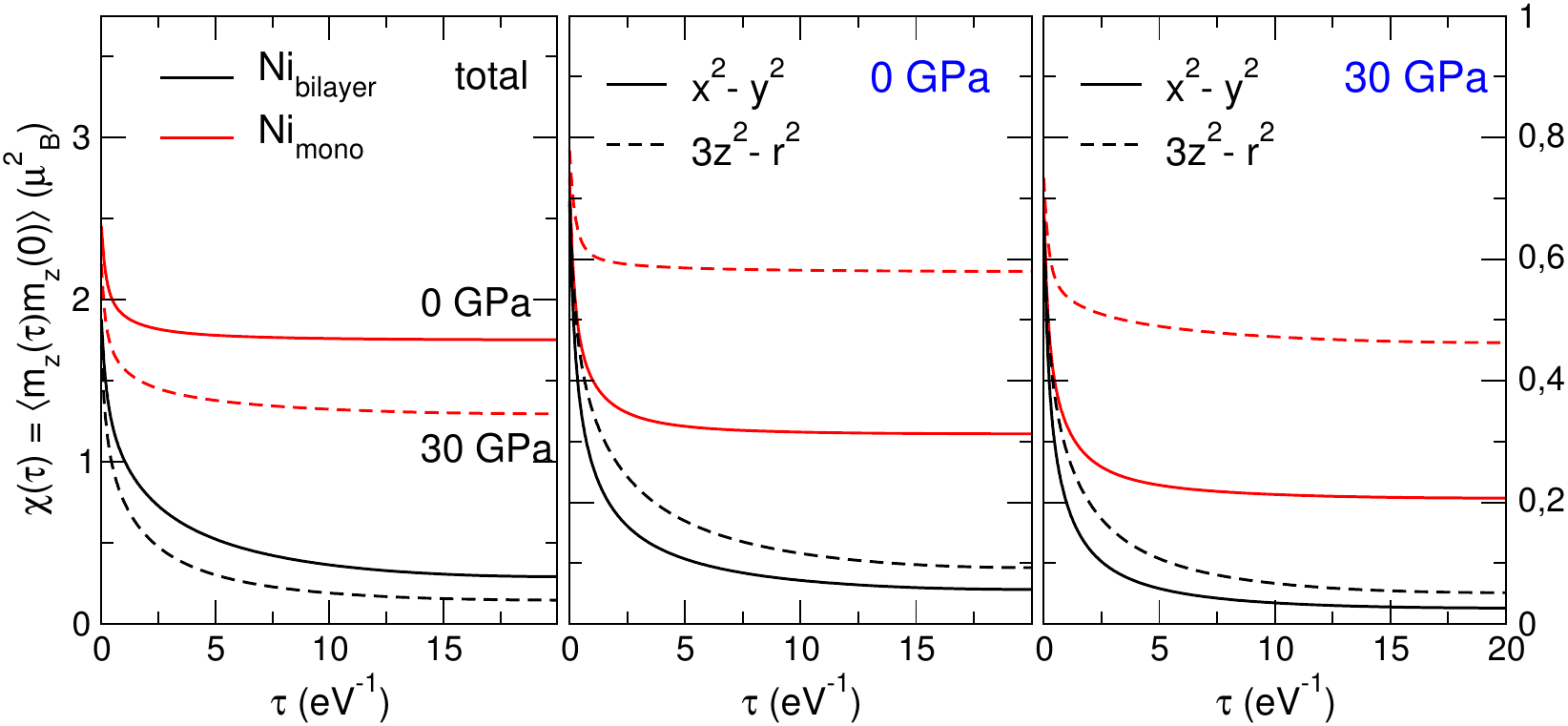}}
\caption{Total and orbitally resolved local spin correlation functions $\chi(\tau) =
\langle \hat{m}_z(\tau) \hat{m}_z(0)\rangle$ as a function of the imaginary time $\tau$ for the Ni $3d$ orbitals calculated by $\mathrm{DFT}+\mathrm{DMFT}$ for the low- and high-pressure (at 30~GPa) phases at $T = 290$~K.
}
\label{Fig_5}
\end{figure}

Our results for the Ni $3d$ self-energies suggest remarkable layer- and orbital-dependent localization of the Ni $e_g$ states caused by strong correlations and structural confinement effects.
This result is corroborated by our analysis of the orbital-dependent local spin susceptibility $\chi(\tau)$ on the imaginary-time domain $\tau$, evaluated within $\mathrm{DFT}+\mathrm{DMFT}$ (see Fig.~\ref{Fig_5}). In particular, the monolayer Ni $e_g$ states are seen to be strongly localized with slowly decaying $\chi(\tau)$ from 0.73/0.78 $\mu_\mathrm{B}^2$ to 0.31/0.58 $\mu_\mathrm{B}^2$ for the $x^2-y^2$/$3z^2-r^2$ orbitals at $\tau=\beta/2$. This gives a relatively large value of the fluctuating moment of about 1.34$\mu_\mathrm{B}$, evaluated as $M_\mathrm{loc} = [k_BT \int \chi(\tau)d\tau]^{1/2}$, where $\chi(\tau)=\langle \hat{m}_z(\tau) \hat{m}_z(0)\rangle$ is the local spin-spin correlation function. The latter is comparable to the instantaneous magnetic moment $\sqrt{ \langle \hat{m}_z^2\rangle}$ of about 1.57$\mu_\mathrm{B}$. We note that the monolayer Ni $x^2-y^2$ orbital states show less localized behavior than the Ni $3z^2-r^2$ ones, being close to the Mott transition (Mott criticality). Overall, these findings are consistent with the orbital-selective Mott insulating solution obtained for the monolayer Ni $e_g$ states. 

Our results for the bilayer Ni $e_g$ states imply significantly less localized behavior with a strong decrease of $\chi(\tau)$ from 0.68/0.69 $\mu_\mathrm{B}^2$ at $\tau=0$ to 0.06/0.09 $\mu_\mathrm{B}^2$ for the $x^2-y^2$/$3z^2-r^2$ orbitals at $\tau=\beta/2$. Our $\mathrm{DFT}+\mathrm{DMFT}$ results for the fluctuating and instantaneous magnetic moments for the bilayer Ni sites are 0.69 and 1.37$\mu_\mathrm{B}$, respectively. This indicates that the effects of electron-electron correlations for the bilayer Ni $e_g$ states exhibit pronounced orbital dependence. Thus, the $3z^2-r^2$ states exhibit more localized behavior than the Ni $x^2-y^2$ ones, in agreement with our estimate of the quasiparticle renormalizations $m^*/m$. This result is also consistent with significantly distinct bare bandwidths obtained within DFT for these bands.

\begin{figure}
\centerline{\includegraphics[width=0.5\textwidth,clip=true]{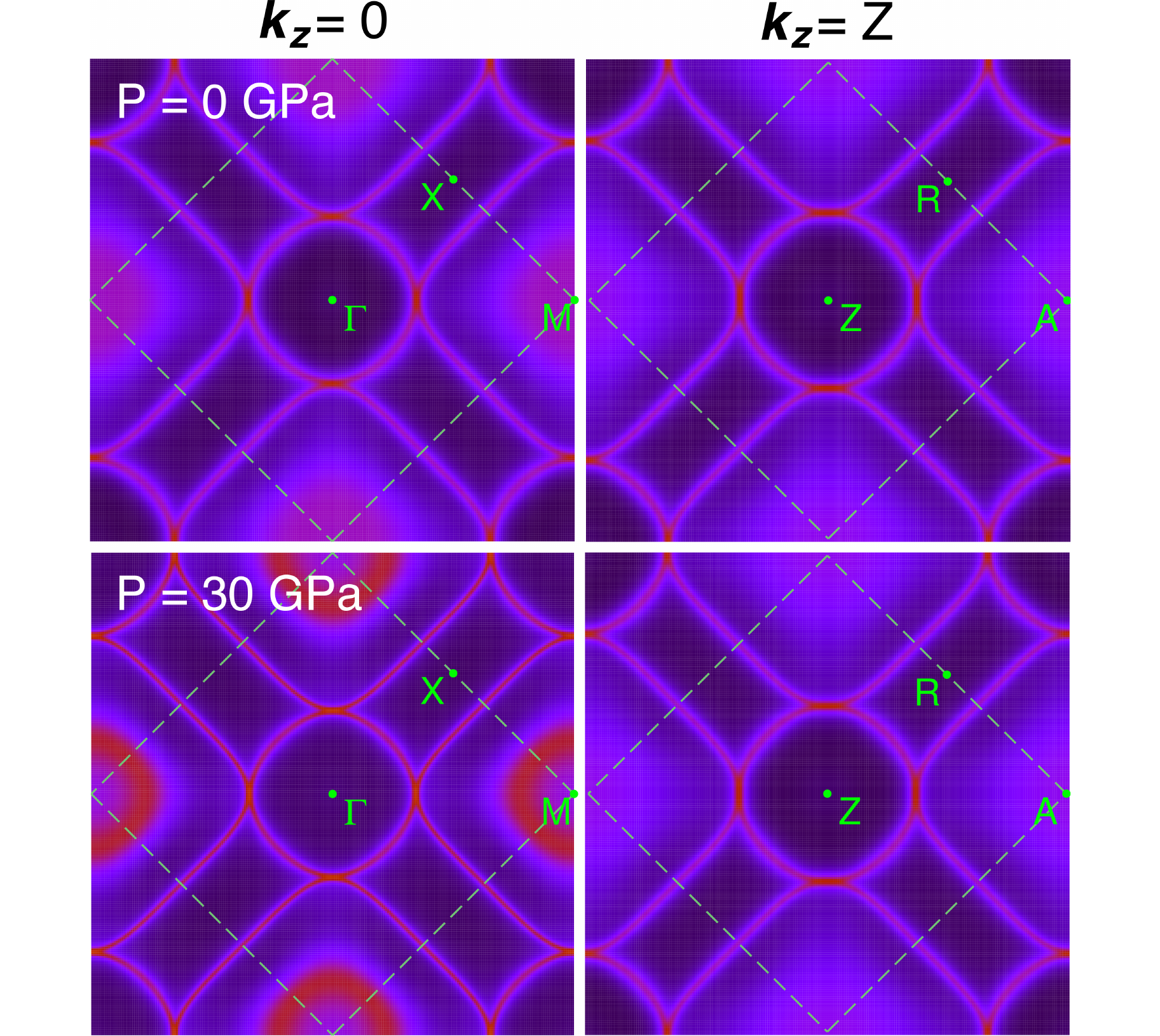}}
\caption{
Correlated Fermi surfaces [spectral function $A({\bf k}, \omega)$ evaluated at $\omega = E_F$] for ${\bf k}_z = 0$ (left) and ${\bf k}_z = \pi$ ($Z$-point, right) calculated within $\mathrm{DFT}+\mathrm{DMFT}$ at $T = 298$~K for different pressures.
}
\label{Fig_6}
\end{figure}

{
\subsection{Fermi surface nesting and magnetic correlations}}

In Fig.~\ref{Fig_6}, we display the in-plane quasiparticle Fermi surface (FS) for different $z$-axis momenta $k_z$ obtained within $\mathrm{DFT}+\mathrm{DMFT}$ at $T=298$~K. The FSs are evaluated as the {\bf k}-resolved spectral function $A({\bf k},\omega)$ at $\omega=0$ for the self-energy $\Sigma(\omega)$ analytically continued on the real energy axis $\omega$. Our results demonstrate a large hole FS sheet centered at the BZ $M$ point which closely resembles that for optimally doped cuprates, including its $x^2-y^2$ orbital character. In addition, we find an electron FS pocket centered at the $\Gamma$ ($Z$) point and a strongly incoherent hole pocket near the $M$ ($A$) point. The latter is associated with a shallow flat band of the (bonding) bilayer Ni $3z^2-r^2$ orbital character. We note that the FSs exhibit strong orbital-dependent incoherence caused by correlation effects. 

Overall, the correlated FSs are two-dimensional and bear a striking resemblance to those of the bilayer nickelate. It is worth noting that correlation effects result in a remarkable reconstruction of the bare (noncorrelated) FS obtained within DFT (see Appendix). In comparison to the DFT result, we observe strong suppression of two square-like FS sheets centered at the BZ corner $M$ and at the $\Gamma$ point (see Fig.~\ref{Fig_S1}). These FS sheets are predominantly of the monolayer Ni $e_g$ orbital character (with the former mainly consisting of the Ni $3z^2-r^2$ orbital character, while the FS pocket centered at $\Gamma$ is associated with the Ni $x^2-y^2$ states). Both FS sheets are suppressed because of strong correlation effects, consistent with the emergence of an orbital-selective Mott state. In contrast, for the FS pockets associated with the bilayer Ni ions, the FSs are notably similar to those obtained within DFT. Thus, the effects of electronic correlations for the bilayer Ni $e_g$ states mainly result in orbital-dependent incoherence of their spectral weight.

\begin{figure}
\centerline{\includegraphics[width=0.5\textwidth,clip=true]{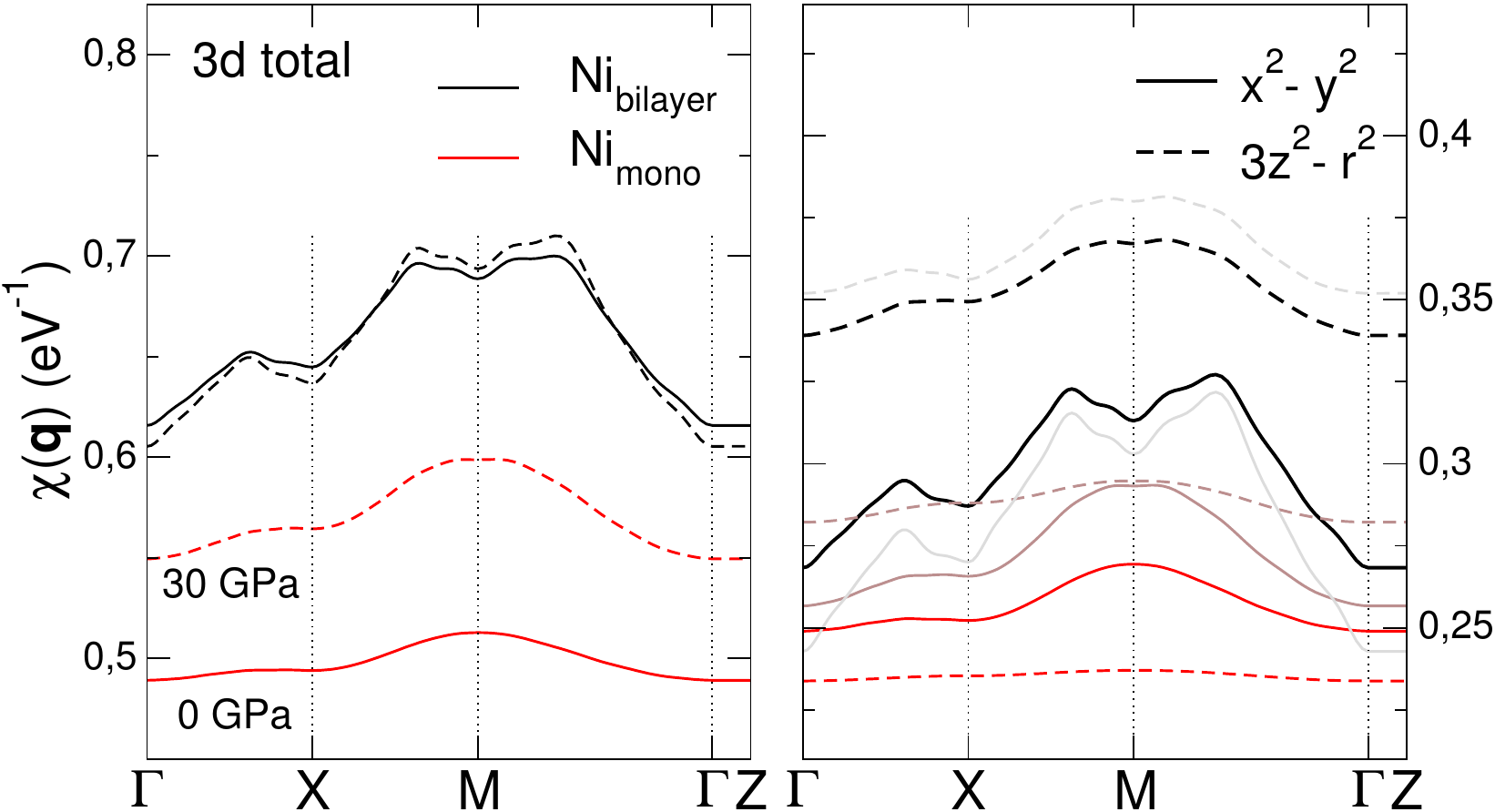}}
\caption{Orbital-dependent in-plane static spin susceptibility $\chi({\bf q})$ of PM 1212-LNO for the crystallographically distinct Ni ions in the bilayer and monolayer NiO$_6$ blocks for different pressures obtained by $\mathrm{DFT}+\mathrm{DMFT}$  at $T = 298$~K using the particle-hole bubble approximation (neglecting  high-order vertex corrections). 
}
\label{Fig_7}
\end{figure}

Our results for the FSs exhibit multiple in-plane nesting effects, implying the emergence of competing long-range density-wave ordering states. We therefore proceed with an analysis of the symmetry and strength of magnetic correlations in 1212-LNO. In this respect, we calculate the momentum-resolved static magnetic susceptibility $\chi({\bf q})$ in the particle-hole bubble approximation  (neglecting the high-order vertex corrections within $\mathrm{DFT}+\mathrm{DMFT}$). It is evaluated as $\chi({\bf q}) = -k_BT \mathrm{Tr} \Sigma_{{\bf k}, i\omega_n}G_{\bf k}(i\omega_n)G_{{\bf k}+{\bf q}}(i\omega_n)e^{i\omega_n0^+}$, where $G_{\bf k}(i\omega_n)$ is the local interacting Green's function for the Ni $3d$ states computed on the Matsubara contour $i\omega_n$ for a given momentum {\bf k}. Our results for different orbital contributions in $\chi({\bf q})$ along the BZ $\Gamma$-X-M-$\Gamma$-Z path are shown in Fig.~\ref{Fig_7}. 

Our $\mathrm{DFT}+\mathrm{DMFT}$ results for $\chi({\bf q})$ exhibit a complex structure with multiple well-defined maxima, suggesting the formation of intertwined spin and charge density wave states in 1212-LNO at low pressure and temperature. We note that magnetic correlations exhibit predominantly the bilayer Ni $e_g$ orbital character with three well-defined maxima located at incommensurate wave vectors on the $\Gamma$-$M$, $X$-$M$, and $\Gamma$-$X$ branches of the BZ. The most pronounced instability is associated with that on the $\Gamma-M$ branch, competing with a minor instability at the $X$-$M$. 

Overall, this result closely resembles that obtained for the bilayer and trilayer nickelates. We note, however, a complex interplay between two double spin-charge density wave states in the bilayer NiO$_6$ block. The former is characterized by the in-plane long-range ordering of the Ni ions moments with a propagating wave vector ${\bf Q} = (\frac{1}{4},\frac{1}{4})$ (with an ``up-up-down-down'' spin configuration pattern), similarly to that in the double-layer nickelate. It strongly competes with long-range  spin-charge density wave ordering with the in-plane wave vector ${\bf Q} = (\frac{1}{3},\frac{1}{3})$ {(e.g., with an ``up-down-0'' spin density wave pattern)}, implying the complex behavior of magnetic correlations in 1212-LNO.
{We note that this behavior can be also interpreted as an instability toward a ``$d^8$--$d^8$--$d^8\underline{L}$'' charge-density wave state with the Ni $d^8\underline{L}$ ions bound in an interbilayer spin polaron, with a hole residing at the O $2p_z$ orbital bridging the two Ni $3d^8\underline{L}$ ions. This behavior is closly similar to a five-spin polaron state proposed recently by J. Li \emph{et al.}~\cite{Li_2026}. 
This suggests an``up-down-UP-up-down-DOWN" spin-density wave ordering, where UP/DOWN are the up/down spins of the Ni $d^8\underline{L}$ ions bonded in the interlayer spin polaron.}

In addition, we observe a minor instability with $\mathrm{max}[\chi({\bf q})]$ at the commensurate wave vector ${\bf q}=(\pi,\pi)$ ($M$ point) associated with the monolayer Ni $e_g$ states. This suggests that the monolayer Ni $3d$ electrons form a long-range N\'eel magnetic state at low temperature. It is interesting that for the monolayer Ni$^{2+}$ ions, both the superexchange and FS nesting effects lead to the same N\'eel-type magnetic state, as obtained within $\mathrm{DFT}+\mathrm{DMFT}$.

It is worth noting that correlation effects result in a remarkable change of the symmetry and strength of magnetic correlations compared to those obtained within DFT. Our DFT results for $\chi({\bf q})$ show that the major instability is associated with the monolayer Ni $e_g$ orbital states (as shown in Fig.~\ref{Fig_S1} these states exhibit square-like FS sheets centered at the BZ $\Gamma$ and $M$ point). We find multiple well-defined maxima of $\chi({\bf q})$ to appear at incommensurate wave vectors on the $\Gamma$-$M$, $X$-$M$, and $\Gamma$-$X$ branches of the BZ, implying competing spin and charge density wave states in the monolayer of 1212-LNO (as obtained within DFT). In contrast, magnetic correlations associated with the bilayer Ni ions are similar to those obtained within $\mathrm{DFT}+\mathrm{DMFT}$. This is associated with the fact that both calculations give similar FSs (and, hence, similar nesting) for the bilayer Ni $e_g$ states, and correlation effects lead only to subtle renormalizations of the spin excitation spectra. 

{
\subsection{Electronic structure and orbital-selective renormalizations under pressure}}

Upon compression, we observe a redistribution of the spectral weights near the Fermi level. It is accompanied by a significant decrease of the quasiparticle mass renormalizations of the bilayer Ni $x^2-y^2$ and $3z^2-r^2$ orbitals, caused by a pressure-driven increase of the bare bandwidth of the Ni $e_g$ states. Indeed, under pressure $\sim$30 GPa the bilayer Ni $x^2-y^2$ and $3z^2-r^2$ exhibit Fermi-liquid-like behavior with the enhancement factors $m^*/m$ reduced by about 30\%, to 2.4 and 3.0 for the Ni $x^2-y^2$ and $3z^2-r^2$ orbitals, respectively. Interestingly, these values are compatible with previously reported results for the double-layer LNO $m^*/m \sim 2.6$ and 3.5 for the Ni $x^2-y^2$ and $3z^2-r^2$ orbitals \cite{Shilenko_2023}. In addition, we notice that the flat-band dispersions associated with the Ni $x^2-y^2$ orbitals in the {\bf k}-resolved spectra do shift away from the Fermi level, to 200 and 400 meV below and above the $E_F$. This behavior is accompanied by a significant pressure-driven shift of the La $5d$ band dispersions toward the Fermi level, which appear above $\sim$0.3 eV. We note that this excludes self-doping characterized by partial occupation of the La $5d$ orbitals at such pressures. 

Most importantly, our results for the $P4/mmm$ 1212-LNO at $\sim$20 GPa (and upon high compression to $\sim$30 GPa) show metallic behavior of the monolayer Ni $x^2-y^2$ and $3z^2-r^2$ orbital states. This implies that under pressure (which is below $\sim$20 GPa) 1212-LNO undergoes an orbital-selective Mott insulator-to-metal phase transition, associated with metallization of the monolayer Ni $e_g$ states. For the monolayer Ni ions the $x^2-y^2$ and $3z^2-r^2$ orbital self-energies show strongly non-Fermi-liquid behavior. The calculated self-energies show no divergence in $\mathrm{Im}[\Sigma(i\omega_n)]$ at the lowest Matsubara frequencies, with a large damping, implying a bad metal behavior characterized by strongly incoherent spectral weights near the Fermi level. The quasiparticle damping evaluated using Pad\'e approximants is 0.89 and 1.4 eV for the $x^2-y^2$ and $3z^2-r^2$ orbital, respectively. 
We propose that under pressure the strongly incoherent single-layer Ni $e_g$ states would contribute to scattering processes resulting in the degradation of superconductivity of the bilayer Ni $3d$ electrons. As a result, this leads to a decrease in the critical temperature in 1212-LNO in comparison to that in the bilayer 2222-LNO \cite{Sadovskii_1997}.
This behavior implies the importance of layer- and orbital-dependent localization of the Ni $e_g$ electrons in 1212-LNO. Overall, the strength of correlation effects is seen to be sufficiently stronger for the monolayer than that for the bilayer Ni $e_g$ orbitals. This highlights the crucial importance of confinement and multiorbital Coulomb correlations to explain the electronic properties of 1212-LNO.

We note that a high compression to about 30 GPa does not affect the $\mathrm{DFT}+\mathrm{DMFT}$ Fermi surface topology, mostly leading to a decrease in the localization of the Ni $e_g$ orbitals. In agreement with this, the fluctuating local moments decrease by about 20\%, to $\sim$1.17 and 0.53$\mu_\mathrm{B}$ for the mono- and bilayer Ni ions, respectively. The calculated instantaneous moments are 1.5 and 1.31$\mu_\mathrm{B}$. Our $\mathrm{DFT}+\mathrm{DMFT}$ calculations of the static spin susceptibility $\chi({\bf q})$ show no qualitative effect of pressure on the spectrum of magnetic excitation. We find that magnetic excitations are predominantly associated with the bilayer Ni $e_g$ orbital states. $\chi({\bf q})$ exhibits essentially the same dependence on the momentum {\bf q} as that at low pressure. As in the ambient pressure results, the most pronounced instability is associated with the $\Gamma-M$ branch, which competes with a minor instability at the $X$-$M$ one. This suggests the formation of intertwined spin and charge density wave stripe states in the NiO$_6$ bilayer of 1212-LNO at low pressure and temperature. Our analysis shows a minor magnetic instability for the monolayer Ni $e_g$ states, with $\mathrm{max}[\chi({\bf q})]$ at ${\bf q}=(\pi,\pi)$ ($M$ point), which is associated with the N\'eel-type magnetic state consistent with our results at low pressure. We note that applied pressure results in a large enhancement in $\mathrm{max}[\chi({\bf q})]$ at ${\bf q}=(\pi,\pi)$ in comparison to the low pressure results, suggesting a possible increase of the corresponding N\'eel temperature under pressure (associated with a crossover from a strong to weak coupling regime). In contrast, magnetic correlations in the bilayer show a modest change under pressure. This suggests that the experimentally observed increase of the spin density wave transition temperature under pressure can be associated with the N\'eel-type magnetic ordering of the monolayer Ni ions. 

\section{Conclusions}

In conclusion, using the $\mathrm{DFT}+\mathrm{DMFT}$ method we explored the effects of electron correlations and pressure on the normal-state electronic structure of the recently synthesized alternating single-layer bilayer hybrid Ruddlesden-Popper nickelate superconductor La$_5$Ni$_3$O$_{11}$ under pressure.
We find that the electronic states originating from the structurally distinct single-layer and bilayer Ni ions show qualitatively different behavior, implying a key role of confinement and orbital-dependent correlations.
For the single-layer Ni ions, we obtain an orbital selective Mott insulator, with a narrow energy gap for the Ni $3z^2-r^2$ states and metallic, strongly incoherent (non-Fermi-liquid) Ni $x^2-y^2$ states. 
In contrast, the $e_g$ states originating from the bilayer Ni ions show strongly renormalized quasiparticle bands with large  enhancement factors $m^*/m \sim 3.5$ and 4.2 for the Ni $x^2-y^2$ and $3z^2-r^2$ states, respectively, implying the proximity of the Ni $e_g$ states to orbital-dependent localization.

We note the emergent flat-band behavior of quasiparticle bands associated with the bilayer Ni $x^2-y^2$ Van Hove states near the Fermi level.
Our analysis of the Fermi surface and magnetic correlations show quantitative similarity to those for the double-layer 2222-LNO.
It shows the possible formation of double spin-charge density wave stripes in the bilayer NiO$_6$ block with a leading instability associated with a propagating wave vector ${\bf Q} = (\frac{1}{3},\frac{1}{3})$.
{This behavior can be understood in terms of two possible scenarios. The first is an instability toward a spin-charge-density wave with an ``up-down-0'' spin pattern. The second is the ``$d^8$--$d^8$--$d^8\underline{L}$'' charge-density wave state, in which the Ni $d^8\underline{L}$ ions are bound in the interbilayer spin polaron with a hole residing in the O $2p_z$ orbital that bridges the two Ni $3d^8\underline{L}$ ions. This implies
the ``up-down-UP-up-down-DOWN" spin-density wave ordering, where UP/DOWN are the up/down spins of the Ni $d^8\underline{L}$ ions bonded in an interlayer spin polaron. Both arrangements possibly compete.
We also note a minor instability associated with} a ${\bf Q} = (\frac{1}{4},\frac{1}{4})$ bicollinear stripe state with  an ``up-up-down-down'' spin pattern similar to that in 2222-LNO, implying a more complex behavior of magnetic correlations in 1212-LNO.
This suggests a key role of intertwined spin and charge density wave stripes in determining the properties of 1212-LNO, in close similarity to the bilayer and trilayer systems.

Under pressure (below 20 GPa), in 1212-LNO an orbital-selective Mott insulator-to-metal phase transition takes place, associated with (hidden) non-Fermi-liquid behavior of the single-layer Ni $e_g$ states.
We show that correlation effects result in a reconstruction of the symmetry and the strength of magnetic correlations compared to those obtained within DFT. Thus, we observe a crossover from single-layer (in DFT) to a double-layer dominated regime.
Overall, our results imply the crucial importance of correlation effects and spin-charge density waves in understanding the unusual properties of 1212-LNO under pressure. This topic calls for further theoretical and experimental research on the complex interplay between charge order, magnetism, and superconductivity established in superconducting nickelates.

%%%%%%%%%%%%%%%%%%%%%
\section{ACKNOWLEDGMENTS}
The DFT electronic structure calculations were supported within the framework of the state assignment of the Ministry of Science and Higher Education of the Russian Federation for the IMP UB RAS. The $\mathrm{DFT}+\mathrm{DMFT}$ calculations, theoretical analysis of the electronic structure and magnetic properties were supported by the Russian Science Foundation (Project No. 25-12-00416) \cite{RSF}.

\section{Appendix: Additional DFT data}

\begin{figure}
\centerline{\includegraphics[width=0.5\textwidth,clip=true]{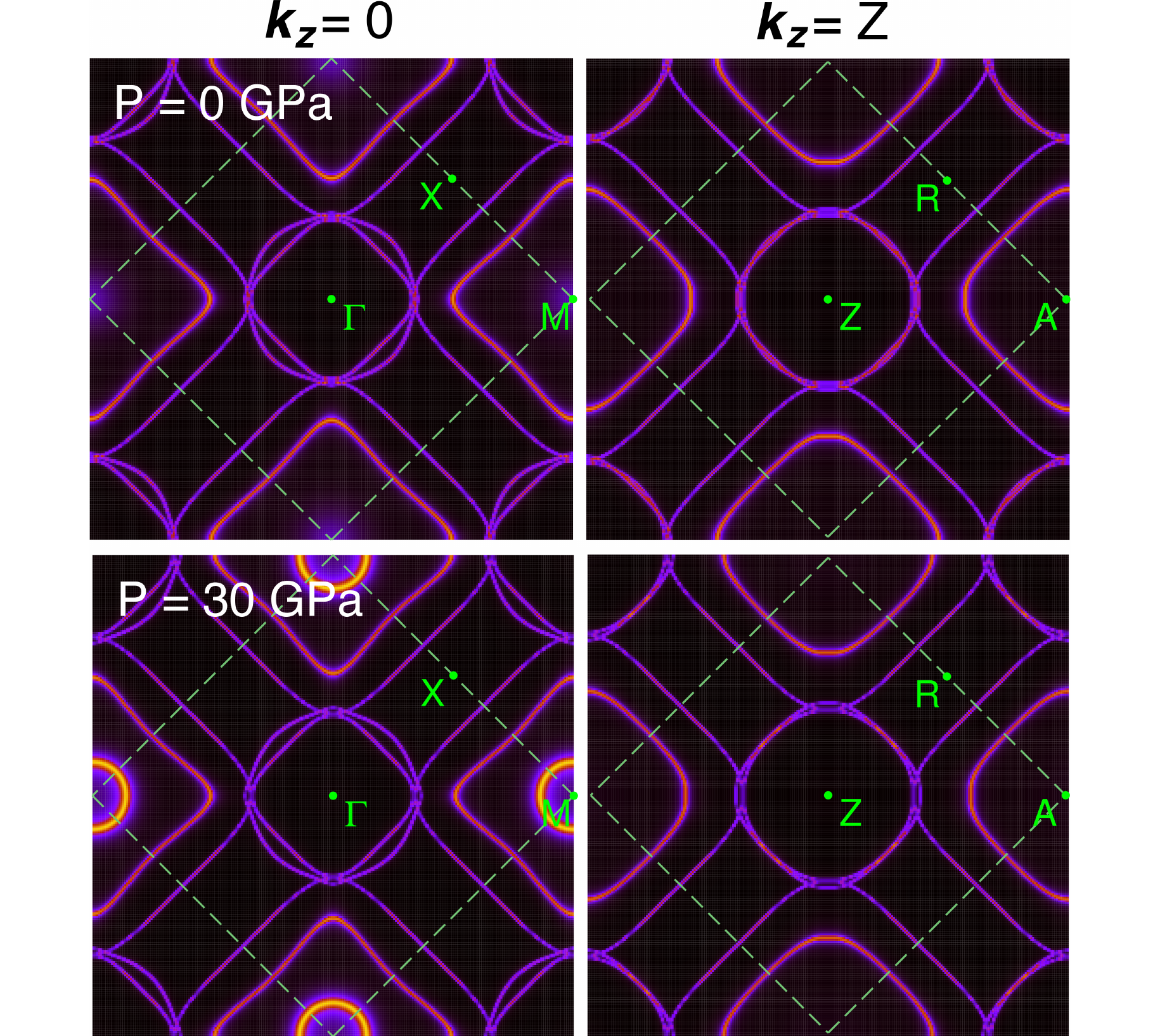}}
\caption{
Fermi surface cuts of 1212-LNO for ${\bf k}_z = 0$ (left) and ${\bf k}_z = \pi$ ($Z$-point, right) calculated using nonmagnetic DFT for different pressures.
}
\label{Fig_S1}
\end{figure}

\begin{figure}
\centerline{\includegraphics[width=0.5\textwidth,clip=true]{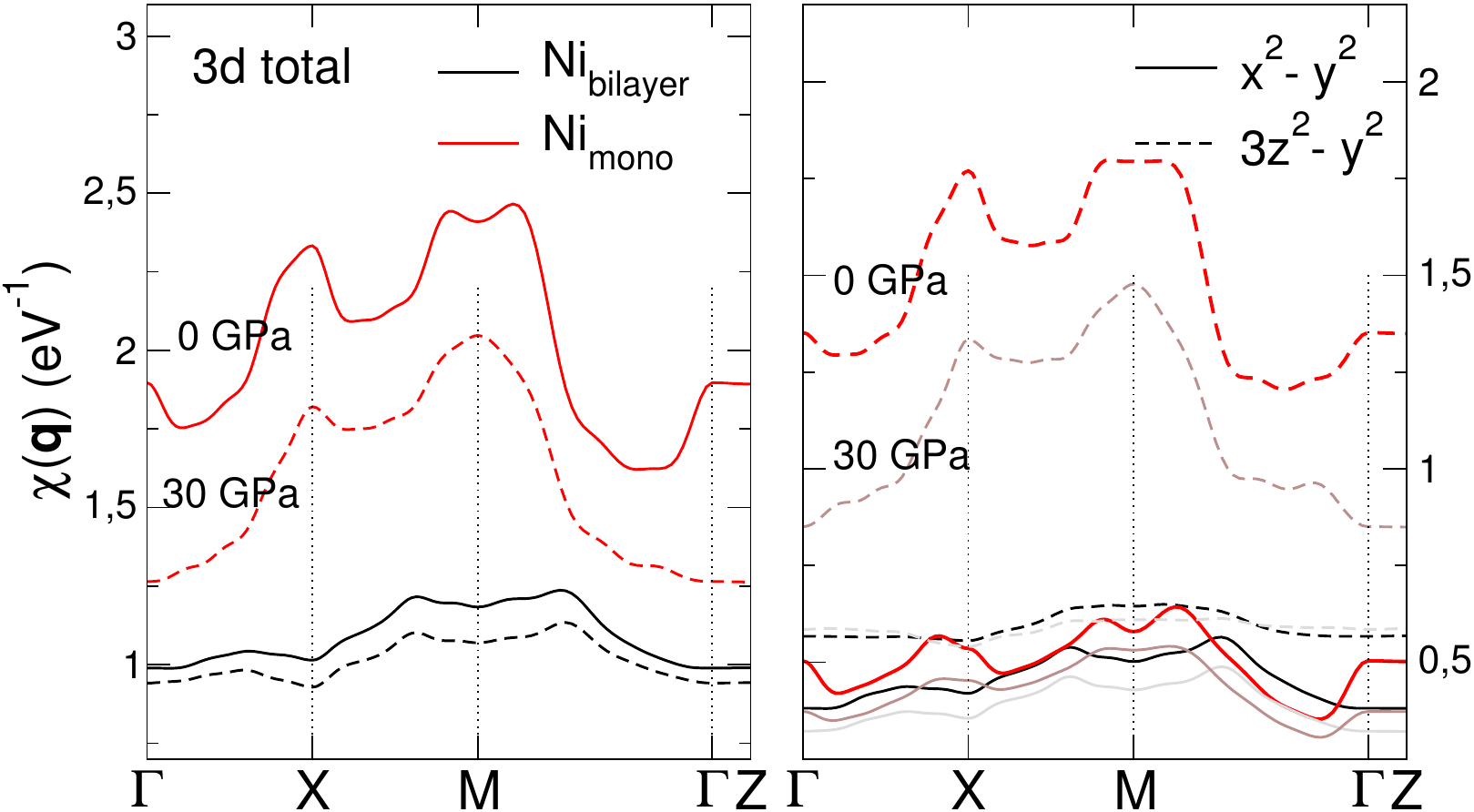}}
\caption{Orbital-dependent in-plane static spin susceptibility $\chi({\bf q})$ of 1212-LNO obtained with the nonmagnetic DFT using the particle-hole bubble approximation (neglecting high-order vertex corrections). 
}
\label{Fig_S2}
\end{figure}

In this appendix, we provide additional data obtained using the nonmagnetic DFT band structure calculations for 1212-LNO at low and high pressures of $\sim$30~GPa. In Figs.~\ref{Fig_S1} and \ref{Fig_S2}, we show our results for the FSs and the corresponding orbital-dependent static spin susceptibility $\chi({\bf q})$ obtained within DFT. The FSs show quasi-two-dimensional behavior with a large hole sheet centered at the BZ $M$ point (which resembles that of optimally doped cuprates) and an electron circle-like pocket centered at the $\Gamma$ ($Z$) point. Moreover, there are two additional square-like FS sheets: one centered at the BZ corner $M$ point and another centered at the $\Gamma$ point (inside of the circle-like FS pocket). These two FS sheets are mainly of the monolayer Ni $e_g$ orbital character, with the former predominantly consisting of the Ni $3z^2-r^2$ orbital character, while the FS pocket centered at the $\Gamma$ point is mainly associated with the Ni $x^2-y^2$ states.

Our DFT results for $\chi({\bf q})$ exhibit multiple well-defined maxima, implying the importance of long-range magnetic correlations characterized by an ordering wave vector driven by Fermi surface nesting effects (see Fig.~\ref{Fig_S2}). In contrast to the $\mathrm{DFT}+\mathrm{DMFT}$ results, the leading instability is associated with the monolayer Ni $e_g$ orbital states, in qualitative agreement with the previous DFT-based calculations \cite{ZhangLin_2025,ZhangChen_2026}. We find that for the monolayer Ni ions $\chi({\bf q})$ there are two major instability channels at the incommensurate wave vectors near the $M$ point, concomitant with a minor instability at the $X$ point. This implies the formation of intertwined spin and charge density wave stripe ordering of the monolayer Ni ions, competing with the striped antiferromagnetic state (an instability at the $X$ point) at low pressure. It is interesting that under a pressure of 20 GPa, it evolves into a commensurate N\'eel-type ordering of the Ni magnetic moments, associated with an instability at the $M$ point.

For the bilayer Ni $e_g$ states, the DFT calculations reveal a less pronounced instability located at an incommensurate wave vector on the $\Gamma$-$M$ branch, competing with a minor instability at the $X$-$M$. Overall, this result closely resembles that obtained for the bilayer and trilayer RP systems. In fact, its momentum and orbital dependence is qualitatively similar to our $\mathrm{DFT}+\mathrm{DMFT}$ results for 1212-LNO. Overall, this implies the possible formation of a long-range density wave stripe ordering in both the single-layer and double-layer NiO$_6$ blocks (at low pressure and temperature), implying the crucial importance of spin fluctuations in explaining the properties of 1212-LNO.

\end{document}